# Evaluation of Software Architecture Quality Attribute for an Internet Banking System

[1] A.Meiappane, [2] B. Chithra, [3] Prasanna Venkataesan, PhD.

[1] Research Scholar, [3] Associate Professor, Pondicherry University,
Puducherry – 605 014.
[2] PG Student, Department of computer science, Sri Manakula Vinayagar Engineering College,
Puducherry

## ABSTRACT

The design phase plays a vital role than all other phases in the software development. Software Architecture has to meet both the functional and non-functional quality requirements.

The Evaluation of Architecture has to be performed, so that the developers are assured that their selected Architecture will reduce the cost and effort and also enhances the various quality attributes like Availability, Reusability, Performance, Modifiability and Extendibility. The success of the system depends upon the Architecture Evaluation by the essential method to the system.

The overall ranking of the candidate architecture is ascertained by assigning weight to the scenario and scenario interaction. In this paper, SAAM method is taken to evaluate the two architectures from the various available method and techniques to achieve the various quality attributes by weight metric.

## 1. INTRODUCTION

Architecture has emerged as a crucial part of the design process. An architecture is the result of a set of business and technical decisions. There are many factors that influence the design of the architecture. An effective technique to assess candidate architecture before implementation and deployment is of great economic value. With the advent of repeatable, structured methods such as Software Architecture Analysis Method (SAAM), architecture evaluation has become a relatively low-cost risk mitigation capability. An architecture evaluation should be a standard part of every architecture based development methodology.

SAAM method is scenario based method which mainly addresses on the modifiability but can be adapted to other attributes. Analyzing the architecture of a software system before acquisition has a large payoff.

Software Architecture Analysis Method (SAAM) is a scenario-based architecture evaluation technique. Analyzing the architecture of a software system using SAAM would guarantee whether the architecture of the system under consideration would achieve the stated goals in terms of functionality and quality attributes. This information can be used to acquire a software system with the guarantee that it would suit the needs of the user.

## 2. SOFTWARE ARCHITECTURE

Software architecture is defined as, "The software architecture of a program or computing system is the structure or structures of the system, which comprise software elements, the externally visible properties of those elements, and the relationships among them"[1].

The architectural view of a system is abstract. It is the overall structure of the system. The architecture of a software system defines the system in terms of computational components and interactions among those components. Components are things such as clients and servers, databases, filters and layers in a hierarchical system[2].

At the architecture level, relevant system-level issues typically include properties such as capacity, throughput, consistency and component compatibility.

### 2.1. Quality Attributes

A quality attribute is a nonfunctional characteristic of a component or a system. A software quality is defined in IEEE 1061 [6] and it represents the degree to which software possesses a desired combination of attributes. Another standard, ISO/IEC Draft 9126-1 [7], defines a software quality model. According to this, there are six categories of characteristics – functionality, reliability, usability, efficiency, maintainability, and portability, which are divided into sub-characteristics. These are defined by means of externally observable features for each software system.

## 3. SOFTWARE ARCHITECTURE EVALUATION

The technique of analyzing whether a given architecture would satisfy the stated goals is not a new one. During recent years, the notion of software architecture has emerged as the appropriate level for dealing with software quality [5].

The architecture of any system directly affects the success of the system. Thus it becomes imperative to validate the architecture of a system as early as possible in the development process, in order to ensure that it is going on the right track.

The SAAM evaluation method is used in the evaluation of Architecture as that is the best method which suites the Architecture Evaluation. This is the first and best method for Software Architecture evaluation.

# 4. EVALUATION METHODS

## 4.1 Software Architecture Analysis Method (SAAM)

Software Architecture Analysis Method (SAAM) [1] is a five step method for analyzing software architectures. The steps or activities are listed below

1. Characterization of functionalities for a given domain.
2. Mapping of this functionalities over the structural decomposition of architecture.
3. Select the quality attributes for assessing the architecture
4. To test the selected quality attribute, identify the concrete tasks.
5. Overall Evaluation of Architecture for these tasks

The main objective of SAAM is Architectural suitability and risk analysis. The direct and indirect scenarios are identified. The number of components affected by scenario is counted by the impact analysis. The objects are analyzed in Architectural documentation especially showing the logical views.

This method accompanied with an architectural description language which was used to analyze three competing architecture with respect to modifiability. This language describes the structural perspective of the competing architectures. It is a practical and proven method and applied for examining architectures of user interface portion of interactive system. It uses three perspectives for understanding and describing architectures which are Functionality, Structure, and Allocation.

SAAM used these three perspectives of architecture description to evaluate system with respect to modifiability. Modifiability was the key quality attribute used in SAAM case study. However, SAAM can be used to evaluate other quality attributes as well.

## 4.2 Scenario-based Software Architecture Analysis method: SAAM

Scenario-based Analysis of Software Architecture [3] shortly (SAAM) is a structured scenario-based architectural analysis. It is the refined form of Software Architecture Analysis Method (SAAM). This version of SAAM has five activities explained below:

### 1. Describe The Candidate Architecture:

This activity emphasizes the architecture analysis with well defined architectural description language that all stakeholders can understand.

### 2. Develop Scenarios:

In this activity the scenarios of tasks for all software system stakeholders are developed.

### 3. Evaluate Each Scenario:

In this activity Direct and Indirect Scenarios are identified with the associated cost and effort.

### 4. Reveal Scenario Interaction:

In this activity scenarios with high interaction are separated from low interaction.

### 5. Weight Scenarios and Scenario Interactions:

In this activity scenarios are weighted and ranked subjectively based on the relative importance of these scenarios with respect to all stakeholders.

$$\text{Wt. of Module} = \text{Module} / \text{Functionality}$$

## 4.4 Software Architecture Analysis Method for Evolution and Reusability: SAAMER

Software Architecture Analysis for Evolution and Reusability [4] (shortly as SAAEMR) is a framework and a set of architectural views designed to evaluate software architecture for evolution and reuse. It is based on SAAM which is a scenario based approach for software evaluation[9]. This framework consists of four phases described below.

**Gathering**

In this phase four different set of information is gathered which includes Stakeholder Information, Architecture Information, Quality Information and Scenarios.

**Modeling**

Modeling is second phase of the framework in which information is aligned across information categories and mapped information into usable artifacts.

**Analyzing**

Analyzing is the third phase of the framework in which SAAM is used for further analysis of various artifacts generated in the last phase.

**Evaluating**

Evaluating is the fourth phase in which recommendations are made, risk and their mitigation strategies are suggested, and common reference models are identified.

The framework presented in this paper can be used for better estimation of cost, schedule, and risk at early stage of software development for evolution and reusability.

The reusability has to be achieved in any software. For evaluation, every scenario is assigned a weight of affected components in the scenario divided by the total number of components in the current architecture (Table 1). For reusability this should be as close to one as possible, i.e., as many of the components as possible should be reusable as-is. Note that for reasons of space the scenarios are presented as *vignettes* [14].

**R1** Visiting the bank web site, those who are interested to view the site or the page can view it.

**R2** Registering into the bank web site, the viewer anyone is interested to be the member or user of the site can also get registered.

**R3** Providing unique user id and password, once the user registered with the bank web site, the unique user id and password has to be generated automatically and provided to the user.

**R4** Entering and Editing your personal details, the user can enter his personal details and also he alone can edit the details entered by him.

**R5** Balance Enquiry, eg, the account holder may need to view the balance of his account up-to-date .

All presented scenarios require behaviour not present in the initial software architecture. However, the scenarios are realistic and measurement systems exist that require functionality defined by the scenarios. The following figure 1, shows the reusability of various scenarios in the software development.

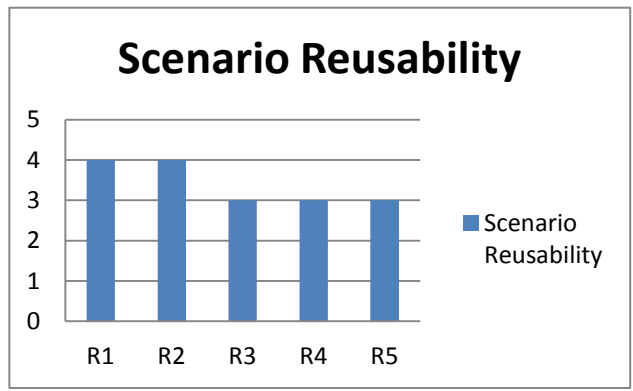

**Figure 1 . Reusabilty of Scenarios**

## 5. RESEARCH ISSUES

The Adaptive Architecture and Pattern oriented Software Architecture are the architecture where the evaluation method has to be reviewed and discussed for internet banking system. There are some scenarios and their various scenario interaction helps to bring out the evaluation in the Business Process Management of Internet Banking System. Therefore it is an innovative essential for the evaluation method with their experimental results particularly to the software quality attributes like Maintainability, Performance, Adaptability and Modifiability/

## 6. CONCLUSION

Since the review was conducted to select appropriate method for software architecture evaluation. The comparative analysis given in     Table - 2 reveals that SAAM provides better support for evaluation. The scope of SAAM is broad and provides integrated capabilities of SAAM, Scenario Based, SAAMER and 4+1 View Model. Organization may use any method/techniques within SAAM. Another conclusion drawn for review is that 4+1 view model  is best for analysis of portion of the architecture. Finally SAAM can also be implemented in small projects and can be tailored [8] for individual software application project which is developed by a small vendor. Therefore, these processes can be integrated in personal software process for individual developer in  a corporate.

## 7. FUTURE WORK

In future it has been planned to review and evaluate the software quality which is not covered in this review. The software quality has to be improved by the various scenario elicitations. The integration of evaluation methods will help in enhancing the individual developers knowledge, skills, and competencies. The integrated process will also help to better understand the impact of architecture design and evaluation on cost, schedule and risk for developers that use personal software process at their stage The evaluation of scenarios are considered in the internet banking system and the same can be carried out for the Mobile Banking quality attributes.

**TABLE - 1: Analysis of Architecture**

| SOFTWARE QUALITY | Scenario | ITERATION NO. | | | | | |
|---|---|---|---|---|---|---|---|
| | | 0 | 1 | 2 | 3 | 4 | 5 |
| REUSABILITY | R1 | 2/4 | 3/5 | 4/6 | 3/9 | 3/9 | **4/10** |
| | R2 | 2/4 | 3/5 | 4/6 | 3/9 | 3/9 | **4/10** |
| | R3 | 0/4 | 0/5 | 1/6 | 2/9 | 2/9 | **3/10** |
| | R4 | 0/4 | 0/5 | 1/6 | 2/9 | 2/9 | **3/10** |
| | R5 | 0/4 | 0/5 | 1/6 | 2/9 | 2/9 | **3/10** |

**TABLE - 2: Comparative Analysis of Software Architecture Evaluation Methods**

| Attributes / Evaluation Method | Availability | Modifiability | Performance | Security | Usability | Reusability |
|---|---|---|---|---|---|---|
| **SAAM** | Y | Y | Y | Y | Y | |
| **4+1VM** | Y | Y | Y | Y | Y | |
| **SAAM (Scenario Based)** | Y | Y | Y | Y | Y | Y |
| **SAAMER** | Y | Y | Y | Y | Y | Y |
| Yes = Y; No = N; Partially Address = P | | | | | | |